\documentclass[twocolumn,prb,superscriptaddress,showpacs]{revtex4}% Physical Review B
\newcommand{\lsim} 
 {\ \raise.35ex\hbox{$<$}\kern-0.75em\lower.5ex\hbox{$\sim$}\ }
\newcommand{\gsim}
 {\ \raise.35ex\hbox{$>$}\kern-0.75em\lower.5ex\hbox{$\sim$}\ }

\newcommand{\mean}[1]{\left<#1\right>}
\newcommand{\means}[1]{\langle#1\rangle}

\def\journal #1#2#3#4{#1 {\bf #2}, #3 (#4)}
\def\PRB{Phys.\ Rev.\ B}
\def\PRL{Phys.\ Rev.\ Lett.}

\def\JMMM{J.~Mag.~Mag.~Mat.}

\def\JPSJ{J.\ Phys.\ Soc.\ Jpn.}

\def\EPL{Europhys.\ Lett.}

\def\NATP{Nat.\ Phys.}
\usepackage{graphicx}
\usepackage{dcolumn}
\usepackage{bm}
\usepackage{amsmath}
\usepackage{times}
\usepackage[usenames]{color}
\usepackage{natbib}

\begin{document}
%\draft
%\preprint{?????}
\title{Ordering and Excitation in Orbital Compass Model on a Checkerboard Lattice}  
\author{Joji~Nasu}
\affiliation{Department  of Physics,  Tohoku University,  Sendai 980-8578,  Japan}
\author{Synge~Todo}
\affiliation{Institute for Solid State Physics, University of Tokyo, 7-1-26-R501 Minatojima-Minamimachi, Chuo-ku, Kobe 650-0047}
\author{Sumio~Ishihara}
\affiliation{Department  of Physics,  Tohoku University,  Sendai 980-8578,  Japan}
\date{\today}
\begin{abstract}  
We study an orbital compass model on a checkerboard lattice where orbital degree of freedom is represented by the pseudo-spin operator. Competition arises from an Ising interaction for the $z$ component of pseudo-spins along the vertical/horizontal bonds and an Ising interaction for the $x$ component along diagonal bonds. Classical and quantum compass models are analyzed by utilizing several analytical methods and numerical simulations.  At a fully frustrated point where the two Ising interactions compete with each other, a macroscopic number of orbital configurations are degenerate in a classical ground state. This degeneracy is lifted by thermal and quantum fluctuations, and a staggered long-range order of the $z$ component of the pseudo-spin is realized. A tricritical point for this order appears due to competition between the bond dependent Ising interactions. Roles of geometrical frustration on excitation dynamics are also examined. 
\end{abstract}

%75.25.Dk 	Orbital, charge, and other orders, including coupling of these orders
%75.30.Et 	Exchange and superexchange interactions (see also 71.70.Gm Exchange interactions)
%75.47.Lx 	Magnetic oxides

\maketitle
\narrowtext

%--- title ---

%--- author ---

%
%
%--- address ---

%
%--- date ---

%   Abstract
%-----------------------------------------------------------

%-----------------------------------------------------------

%\section{Introduction\label{sec:intro}}
%%%%%%%%%%%%%%%%%%%%%%%%%%%%%%%%%%%%%%%%%%

\section{Introduction}
Long-range order and excitation dynamics in the orbital degenerate correlated electron systems are one of the recent attractive themes in condensed matter physics.~\cite{book} Orbital degree of freedom represents a spatially anisotropy of the electronic wave function. In molecules, orbital degeneracy is usually lifted by coupling with lattice, i.e. the Jahn-Teller effect. In contrast, in crystal lattices, there are some equivalent bonds around a transition-metal ion. When an orbital is directed along one of the equivalent bonds, anisotropy in the bond energies comes out. In this sense, all bond energies on the equivalent bonds are not minimized simultaneously. This is regarded as a kind of frustration effect termed ``orbital frustration''.~\cite{khomskii03,Ishihara97} This characteristic in the orbital degenerate systems provides a wide variety of exotic phenomena such as order by disorder  phenomena,~\cite{Kubo,Nussinov_EPL,Khaliullin_Okamoto} orbital liquid state~\cite{Feiner,Khaliullin_Maekawa} and so on. 

One of the well-studied orbital models is the Kugel-Khomskii model.~\cite{Kugel} This is applied to orbitally degenerate Mott insulators, where the orbital degree of freedom in a transition-metal ion is described by the pseudo-spin (PS) operator. The intersite interactions between the nearest neighbor ions are represented by products of the Heisenberg-type interaction between spins and the orbital interaction. In the orbital part, the interaction between the pseudo-spins explicitly depends on the bond direction. 

Another well studied orbital model is the orbital compass model where the orbital degree of freedom is only taken into account. A general expression of the orbital compass model is given by 
\begin{align}
 H&=J\sum_{<ij>_l}\left(\hat{n}_l\cdot\bm{T}_i\right)\left(\hat{n}_l\cdot\bm{T}_j\right) , 
\label{eq:10}
\end{align}
where $\bm{T}_i$ is the PS operator for the doubly-degenerate orbital degree of freedom with an amplitude of $1/2$, $\hat{n}_l$ is a unit vector along the bond direction $l$, and $\langle ij \rangle_l $ indicates the nearest neighboring (NN) $i$ and $j$ sites along $l$. This model has some analogy to the dipole-dipole interaction and shows the same characteristics with the orbital part in the Kugel-Kohmskii model; the interactions explicitly depend on a bond direction. This model has been studied from broad view points; quantum phase transition,~\cite{Orus} topological quantum order~\cite{Nussinov}, hidden dimer order,~\cite{Brz07} and protected qubit~\cite{Doucot,Gladchenko} are examined on a square-lattice compass model, and a kind of compass model is proposed as an appropriate model for transition-metal oxides with a strong spin-orbit coupling.~\cite{Jackeli} 

Transition-metal ions with orbital degree of freedom sometime consist geometrically frustrated lattices, such as triangle and spinel crystals. Interplay of geometrical frustration and spin-orbital entanglement often give rise to novel states of matter, such as spin-orbital molecules in AlV$_2$O$_4$,~\cite{katsufuji} a cooperative release of frustration proposed in ZnV$_2$O$_4$,~\cite{motome04} and a resonating valence bond state predicted in LiNiO$_2$.~\cite{vernay04} 
Even without spin-orbital entanglement, ground state and excitation dynamics in orbital degenerate system with geometrically frustrated lattice are non-trivial because of the orbital anisotropy and frustration characteristics. The recent neutron scattering experiments suggest an excitation from spin-orbital molecules in GeCo$_2$O$_4$,\cite{tomiyasu} where effective total angular moments might be described by the orbital compass model. 

In this paper, the orbital compass model on one of the geometrically frustrated lattices, i.e. a checkerboard lattice, is studied. Competitions arise from an Ising interaction for the $z$ component of PS, $T^z$, along the horizontal and vertical directions on a lattice and an Ising interaction for $T^x$ along the diagonal directions. Phase diagrams in classical and quantum models, where PS operators are regarded as classical vectors and quantum operators, respectively, are obtained by several analytical and numerical methods. It is shown that, at a fully frustrated point, where a number of classical PS configurations are degenerate, two-dimensional staggered $T^z$ ordered state is stabilized by thermal and quantum fluctuations. Because of the bond depend Ising interactions, a tricritical point for the two-dimensional staggered $T^z$ order appears. A one-dimensional characteristic excitation in this ordered state is remarkable near the phase boundary. Present results are compared with the results in the compass model on a square lattice. 

In Sect.~II, an orbital compass model on a checkerboard lattice is introduced. In Sect.~III, a classical model is analyzed by the mean-field (MF) approximation and the classical Monte-Carlo (MC) simulation. In Sect.~IV, a quantum model is analyzed by the spin-wave approximation, a combined method of the MF approximation and the Jordan-Wigner transformation, the exact-diagonalization method and the quantum MC simulation. In Sect.~V, results for the excitation dynamics are presented. Section VI is devoted to summary and discussion.

\section{Model}

\begin{figure}[!b]
\begin{center}
\includegraphics[scale=0.2]{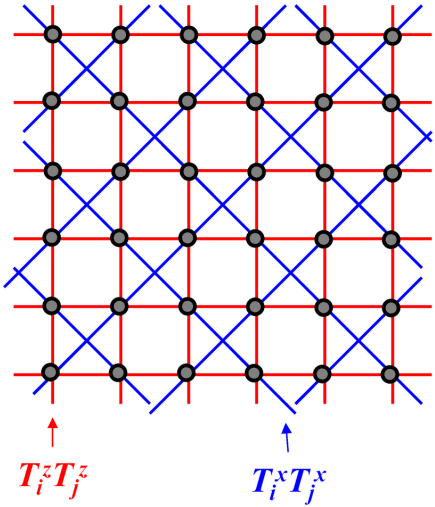} 
\caption{A schematic picture for the orbital compass model on a checkerboard lattice.} 
\label{check}
\end{center}
\end{figure}
We set up an orbital-compass model on a two-dimensional checkerboard lattice. Doubly-degenerate orbitals, $d_{zx}(\equiv a)$ and $d_{yz}(\equiv b)$, are introduced in each site. We focus on the orbital degree of freedom of electrons and neglect the spin degree of freedom. We start from the spinless Hubbard model with the doubly degenerate orbitals defined by 
\begin{align}
 H&=\sum_{<ij>_l}t_{NN;\gamma\gamma'}^{(l)}c_{i\gamma}^\dagger c_{j\gamma'}
+\sum_{<ij>'_m}t_{NNN;\gamma\gamma'}^{(m)}c_{i\gamma}^\dagger c_{j\gamma'}\nonumber\\
&\ \ \ \ \ \ \ \ \ +U\sum_i n_{ia}n_{ib} , 
\end{align}
where  $c_{i\sigma}$ is the annihilation operator for a spinless fermion with orbital $\gamma(=a,b)$  at site $i$, and symbols $<\!ij\!>_l$  and $<\!ij\!>'_m$  represent the  NN- and  the next NN (NNN) $ij$ pairs,  respectively. The transfer integral $t_{NN;\gamma\gamma'}^{(l)}$  ($t_{NNN;\gamma\gamma'}^{(m)}$) is defined on a NN (NNN) bond along the direction $l$ ($m$). Matrix elements of the transfer integrals are determined by the Slater-Koster parameters. Since an electron in the $d_{zx}$ ($d_{yz}$) orbital hops along the $x\ (y)$ direction, the matrix elements for $t_{NN;\gamma\gamma'}^{(l)}$ are given by~\cite{Clay10}
\begin{align}
 \hat{t}_{NN}^{(x)}=
\begin{pmatrix}
-t_1&0\\
0&0
\end{pmatrix}=-\frac{t_1}{2}(1+\sigma^z) ,
\label{eq:1}
\end{align}
and 
\begin{align}
\hat{t}_{NN}^{(y)}=
\begin{pmatrix}
0&0\\ 0&-t_1 \end{pmatrix}=-\frac{t_{1}}{2}(1-\sigma^z) , 
\label{eq:2} \end{align}
with a positive constant $t_1$, where $\bm{\sigma}$ are the Pauli matrices. 
Matrix elements for $t_{NNN;\gamma\gamma'}^{(m)}$ are obtained by introducing the linear combinations of the $d_{zx}$ and $d_{yz}$ orbitals,  i.e. 
$ (d_{zx} \pm d_{yz})/\sqrt{2}$, as 
\begin{align}
 \hat{t}_{NNN}^{(xy)}&=-\frac{t_2}{2}
\begin{pmatrix}
1&1\\
1&1
\end{pmatrix}=-\frac{t_2}{2}(1+\sigma^x) , 
\end{align}
and 
\begin{align}
\hat{t}_{NNN}^{(x\bar{y})}&=-\frac{t_2}{2}
\begin{pmatrix}
1&-1\\ -1&1 \end{pmatrix}=-\frac{t_2}{2}(1-\sigma^x), 
\end{align}
with a positive constant $t_2$. From this Hubbard-type Hamiltonian, an effective Hamiltonian in the case of $U \gg t_1, t_2$ is derived by the second-order perturbational procedure as 
\begin{align}
 H=J_z\sum_{<ij>}T_i^z T_j^z+J_x\sum_{<ij>'}T_i^x T_j^x\label{eq:32},
\end{align}
where we define the exchange constants $J_{z}=2t_{1}^2/U$ and $J_{x}=2t_{2}^2/U$. We introduce the PS operator, ${\bm T}_i$,  with a magnitude of  1/2, where the $d_{zx}$ and  $d_{yz}$ orbitals are taken to be the eigen states of $T^z_i$. This Hamiltonian is a kind of the orbital compass model defined on a checkerboard lattice in a sense that the Ising-type interactions depend on bond directions.

Next we discuss a symmetry of the Hamiltonian in Eq.~(\ref{eq:32}). Let us focus on a NNN bond network on a checkerboard lattice (see Fig.~\ref{check}). One dimensional chains along $\langle 11 \rangle$ and $\langle 1\bar 1 \rangle$ directions are independent with each other. In one of the chains, termed $l$, we introduce the operator defined by~\cite{Doucot}
\begin{align}
 P_l=\prod_{i\in l}\sigma_i^z , 
\label{eq:pl}
\end{align}
where $i$ runs along this chain. It is shown that this operator commutes with the Hamiltonian by using the commutation relation $[\sigma_i^z \sigma_j^z,\sigma_i^x   \sigma_j^x]=0$ with $i \ne j$. Therefore, the energy eigenstates are labeled by the eigenvalues of $P_l$, i.e. $\pm 1$. There are $L$ labels on a $L \times L$-site lattice. This characteristic is available in numerical exact-diagonalization calculations for large cluster size. Because of these local symmetries, the generalized Elitzur's theorem is applicable to this model.~\cite{Batista_Nussinov} It is rigorously shown that a long range order of $T^x_{\bm q} \equiv N^{-1} \sum_i  T^x_i  e^{i\bm{q}\cdot \bm{r}_i}$ for any momenta of ${\bm q}$, which does not commute with $P_l$, is not realized at finite temperature.

\section{Classical Orbital State}  
In this section, we treat the orbital PS  as a classical vector defined in a two-dimensional $T^x$-$T^z$ plane with an amplitude of 1/2.

\subsection{Mean-Field Analysis}  
\begin{figure}[!t]
\begin{center}
\includegraphics[scale=0.5]{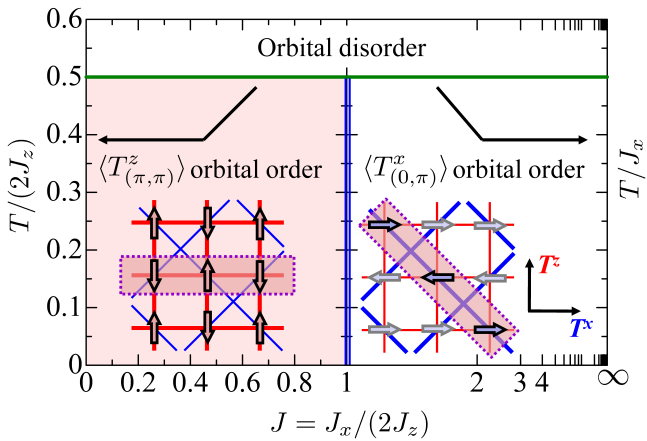}
\caption{
Phase diagram obtained by the MF approximation. The lines in $J<1$ and $J>1$ are plotted in the different scales at left and right figure, respectively. Stable orders are $\langle T^z_{(\pi, \pi)} \rangle$ for $J<1$, and $\langle T^x_{(\pi, 0)} \rangle$ or $\langle T^x_{(0, \pi)} \rangle$ for $J>1$. Transition temperatures do not depend on $J$ and are $T_c/(2J_z)=0.5$ in $J<1$ and $T_c/J_x=0.5$ in $J>1$. Insets show schematic PS configurations in the $T^x-T^z$ plane.
} 
\label{check_mean} \end{center}
\end{figure}
First, we show the orbital state obtained by the MF approximation. We take the MFs for the orbital order as $\langle T^l_{\bm{q}} \rangle=N^{-1} \sum_i \langle T^l_i \rangle e^{i\bm{q}\cdot\bm{r}_i}$ for $(l=x, z)$. In Fig.~\ref{check_mean}, the phase diagram in the plane of $J \equiv J_x/(2J_z)$ and temperature, $T$, is presented. Stable orbital orders are $\langle T^z_{(\pi, \pi)} \rangle$ for $J<1$, and $\langle T^x_{(\pi, 0)} \rangle$ or $\langle T^x_{(0, \pi)} \rangle$ for $J>1$, i.e. the staggered $T^x$ order along $\langle 11 \rangle$ and $ \langle {\bar 1}1 \rangle$ directions. The transition temperatures do not depend on a magnitude of $J$. 
Beyond the analyses for the MF order parameters with single momentum, there are a number of degenerate MF solutions for $J>1$; in the $\langle T^z_{(0, \pi)} \rangle$ ordered state, we consider the transformation of PS that $T^x_i  \rightarrow - T^x_i$ for all sites in a certain chain along $\langle 11 \rangle$ and $\langle {\bar 1}1 \rangle $ directions. The MF energy is not changed under this transformation, since $T^z$ operator is only concerned in the interaction along the $\langle 10\rangle $ and $\langle 01 \rangle $ directions. There are $2^{2L}$ degenerate MF solutions on a $L \times L$-site lattice at $T=0$. 

At a point of $J=1$ and $T=0$, there is an additional degeneracy. Any linear combinations of $\langle T^x_{(0, \pi)} \rangle$ and $\langle T^z_{(\pi, \pi)} \rangle$, i.e. $\langle T(\theta) \rangle=\cos\theta\langle T^z_{(\pi, \pi)} \rangle+\sin\theta\langle T^x_{(0, \pi)} \rangle$ where $\theta$ is the rotation angle in the $T^z-T^x$ plane, have the same energy. This degeneracy is not expected from the Hamiltonian which does not show any continuous symmetry.

\subsection{Monte Carlo Simulation}
In this subsection, we introduce the numerical results obtained by the classical MC simulations. Two-dimensional $20^2$-, $30^2$- and $40^2$-site clusters with a periodic boundary condition are used. We adopt the Wang-Landau algorithm,~\cite{WL}  where $5 \times 10^7$MC steps are used for both making histograms and measurements. 

\begin{figure}[!t]
\begin{center}
\includegraphics[scale=0.45]{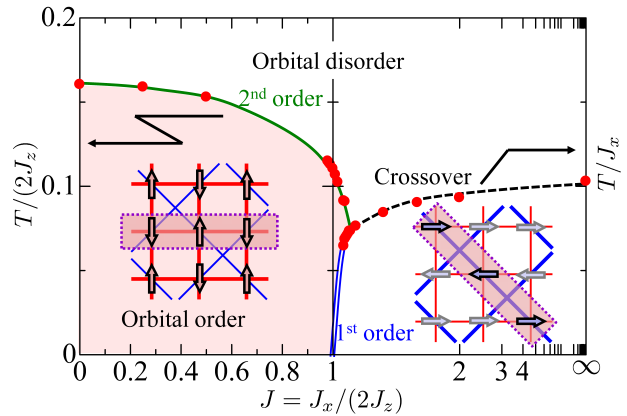}
\caption{
Phase diagram obtained by the classical MC method. The data in $J<1$ and $J>1$ are plotted in the different scales at left and right figure, respectively. Bold and double lines represent the second- and first-order phase transitions, respectively. Dotted line represents the crossover below which an one-dimensional $T^x$ correlation develops. The filled circle at $J=\infty$ is determined by applying the MC simulation to the one-dimensional Ising model on $L=200$ chain. The first order occurs at $J=1$ and zero temperature.
}
\label{check_classical_phase}
\end{center}
\end{figure}
The phase diagram is presented in Fig.~\ref{check_classical_phase}. With increasing $J$ from $J=0$, where the model is reduced to the $T^z$-Ising model on a square-lattice, the transition temperature for the $\langle T^z_{(\pi, \pi)} \rangle$ order gradually decreases because of the competition between $J_z$ and $J_x$. At another limit, $J=\infty$, the model is reduced to the independent one-dimensional $T^x$-Ising model which does not show a long-range order. However, there is a crossover temperature around $T/J_x=0.1$ where the specific heat $C$ shows a broad peak below which an one-dimensional $T^x$ correlation develops. By introducing the NN interaction, $J^z$, the broad peaks remain in the temperature dependences of the specific heat. The peak positions are plotted by dotted lines in Fig.~\ref{check_classical_phase}. The crossover temperature gradually decreases, when the system approaches to the $J=1$ point.
Let us focus on the point of $J=1$. At $T=0$, continuous degeneracy exists as explained above. With increasing $T$, the $\langle T^z_{(\pi, \pi)} \rangle$ order is stabilized among them due to the thermal effect. This is a kind of order by fluctuation phenomena. When temperature is increased furthermore, a conventional thermal effect makes a system to be an orbital disordered state. As a result, reentrant feature is observed in the phase boundary around $J=1$. 

\begin{figure}[!t]
\begin{center}
\includegraphics[scale=0.55]{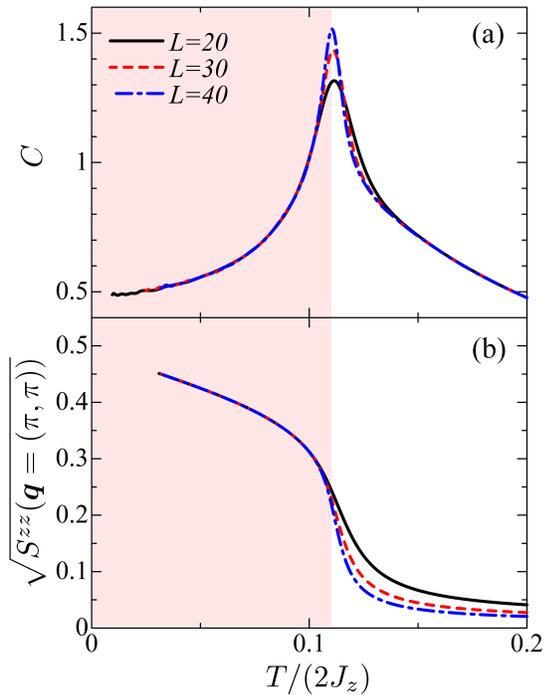}
\caption{
Temperature dependences of (a) specific heat and (b) square root of the staggered-type correlation function $S^{zz}(\pi, \pi)$ for several cluster sizes at $J=1$.}
\label{check_mc2.0}
\end{center}
\end{figure}
Next, we present detailed MC results around $J=1$. In Fig.~\ref{check_mc2.0}, the temperature dependences of the specific heat and the squre root of the staggered PS correlation function $S^{zz}(\pi,\pi)$ are presented on several cluster sizes at $J=1$. 
We define $S^{zz}(\bm{q})=N^{-2} \sum_{ij}T^z_i T^z_je^{i\bm{q}\cdot (\bm{r}_i-\bm{r}_j)}$. A sharp peak is observed in $C$ at $T/(2J_x) \sim 0.12$ which is termed $T_c$ from now on. A value of $C=0.5$ in the limit of $T =0$ implies an existence of one degree of freedom per site, i.e. the polar angle of PS in the $T^x-T^z$ plane. The correlation function starts to increase around $T_c$ and approaches to the upper limit of 0.5 at low temperatures. With increasing the system size, $T_c$ slightly decreases, a peak in $C$ becomes sharp, and an increase in $S^{zz}(\pi, \pi)$ at $T_c$ becomes sharp. The results imply that these anomalies at $T_c$ correspond to the second-order phase transition in the thermodynamic limit.

\begin{figure}[!t]
\begin{center}
\includegraphics[scale=0.65]{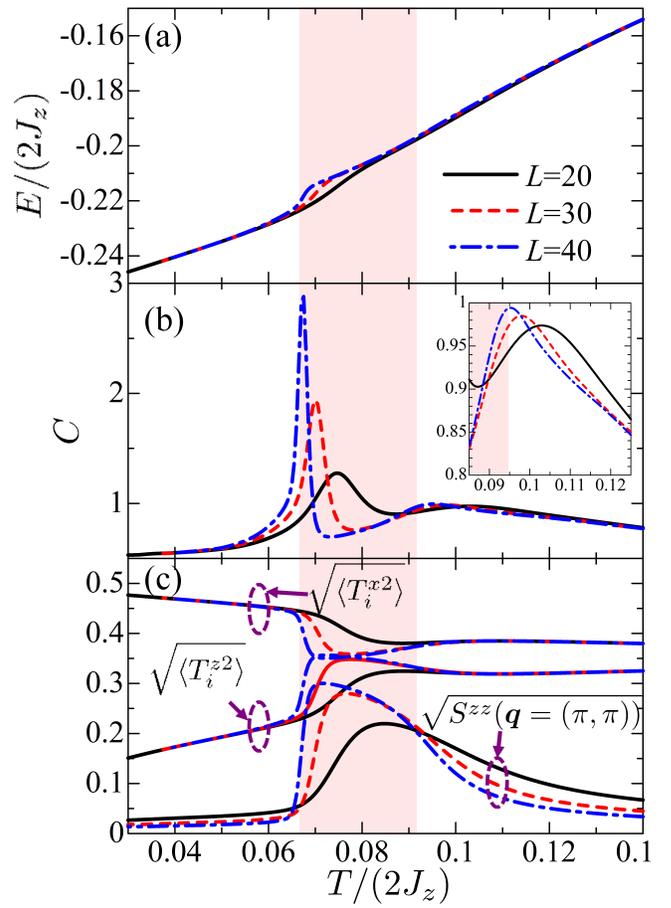}
\caption{
Temperature dependences of (a) energy, (b) specific heat, and (c) square root of the staggered-type correlation function $S^{zz}(\bm{q}=\vec{\pi})$,  root mean squares of the PS moment $\mean{T_i^{z2}}$, and $\mean{T_i^{x2}}$ for several cluster sizes at $J=1.045$.
}
\label{check_mc2.09}
\end{center}
\end{figure}
In Fig.~\ref{check_mc2.09}, the numerical results at $J=1.045$ are presented. We show the temperature dependences of energy $E$, $C$, $[S^{zz}(\pi, \pi)]^{1/2}$ and $\mean{T_i^{l2}}^{1/2}\ (l=x, z)$  for several size clusters. Two anomalies are observed at $T/(2J_z) \sim 0.067$ and 0.09 which are termed $T_L$ and $T_H$, respectively. 
At $T_H$, a peak in $C$ becomes sharp, and an increase in $S^{zz}(\pi, \pi)$ becomes remarkable with increasing the system size. These results are similar to the results at $T_c$ in $J=1$ (see Fig.~\ref{check_mc2.0}). On the other hand, at $T_L$, $S^{zz}(\pi, \pi)$ and $\langle T^{z2} \rangle$ decrease and $\langle T^{x2} \rangle$ increases with decreasing $T$. Large system-size dependences are observed in $E$ and $C$ at $T_L$. 
These results imply that, in the region of $T_L < T < T_H$, the $\langle T^z_{(\pi, \pi)} \rangle$ order is realized in the thermodynamic limit. Below $T_L$, PS's are directed along the $T^x$ axis. This is expected from the one-dimensional $T^x$ Ising interaction. 

\begin{figure}[!t]
\begin{center}
\includegraphics[scale=0.6]{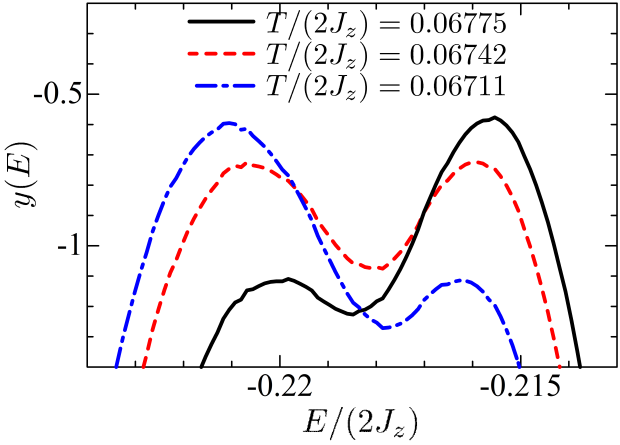}
\caption{Energy dependence of $y(E)=\ln D(E)-\beta EN+{\rm const.}$ (see text) for several temperatures at $J=1.045$. The cluster size is chosen to be $N=40^2$.}
\label{check_mc2.09_hist}
\end{center}
\end{figure}
To examine the orders of the phase transitions at $T_L$, we calculate the free energy defined by 
\begin{align}
F&=-T\ln \sum_E D(E)e^{-\beta EN}+{\rm const.}\nonumber\\
&=-T\ln \sum_E \exp\left[ \ln D(E)-\beta EN +{\rm const.}\right] , 
\label{eq:3}
\end{align}
where $D(E)$ is the density of states. We calculate $y(E) \equiv \ln D(E)-\beta EN +{\rm  const.}$ as the energy histogram in the Wang-Landau scheme in the MC simulation.\cite{WL} The results for three temperatures around $T_L$ are shown in Fig.~\ref{check_mc2.09_hist} as functions of $E$. Double peak structures are commonly observed in $y(E)$. The two peak heights are reversed by changing temperature; $y(E)$ at the lower-energy peak increases with decreasing $T$. Since $y(E)$ is proportional to the system size as expected from the definition, the two minima in $-y(E)$ are separated by energy of the order of $N$. It is expected in the thermodynamic limit that the energy in the stable state is changed discontinuously by changing $T$, and the anomaly at $T_L$ corresponds to the first-order phase transition. 

\section{Quantum Orbital State}  
In this section, the orbital PS's are treated as quantum spin operators with an magnitude of 1/2.

\subsection{Spin-Wave Approximation}
\begin{figure}[t]
\begin{center}
\includegraphics[scale=0.5]{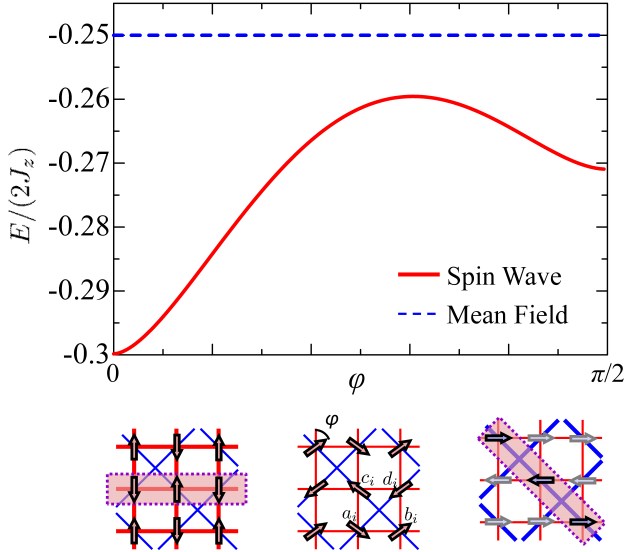}
\caption{
Ground state energy at $J=1$ obtained by the spin-wave approximation (solid line). Broken line represents the MF energy. Schematic PS configurations assumed in the spin-wave approximation are also shown.
}
\label{check_sw_phi}
\end{center}
\end{figure}

The ground state at $T=0$ is analyzed by using the spin wave approximation. The long-range ordered states of $\langle T^z_{(\pi, \pi)} \rangle$ and $\langle T^x_{(0, \pi)} \rangle$ are adopted as the ground states for $J<1$ and $J>1$, respectively, although a number of degenerate states exist for $J>1$. At $J=1$, continuous PS configurations connecting $\langle T^z_{(\pi, \pi)} \rangle$ to $\langle T^x_{(0, \pi)} \rangle$, i.e. $\langle T(\varphi) \rangle=\cos\varphi\langle T^z_{(\pi, \pi)} \rangle+\sin\varphi\langle T^x_{(0, \pi)} \rangle$ are assumed (see Fig.~\ref{check_sw_phi}). There are four sublattices in the ordered states.
 
By introducing the four kinds of the Holstein-Primakoff bosons, $(a_{\bm{k}}, b_{\bm{k}}, c_{\bm{k}}$, $d_{\bm{k}})$, the Hamiltonian up to the second order of the boson operators is given as
\begin{align}
 H_{\rm SW}/(2J_z)&=-NS^2(\cos^2\varphi+J\sin^2\varphi)\nonumber\\
&+S\sum_{\bm{k}}^{N/4} \biggl[2\left(\cos^2\varphi+J\sin^2\varphi\right)h_{\bm{k}}^0 \nonumber \\
&\ \ \ \ \ +\frac{1}{2}h_{\bm{k}}^z\sin^2\varphi +\frac{J}{2}h_{\bm{k}}^x\cos^2\varphi
\biggr],\label{eq:4}
\end{align}
where $S=1/2$, and 
\begin{align}
 h_{\bm{k}}^0&=\bigl(a_{\bm{k}}^\dagger a_{\bm{k}}+b_{\bm{k}}^\dagger b_{\bm{k}}+
c_{\bm{k}}^\dagger c_{\bm{k}}+
d_{\bm{k}}^\dagger d_{\bm{k}}\bigr) , \\
 h_{\bm{k}}^z&=\cos k_x \bigl(a_{\bm{k}}^\dagger b_{\bm{k}}
+c_{\bm{k}}^\dagger d_{\bm{k}}
+a_{\bm{k}}^\dagger b_{-\bm{k}}^\dagger
+c_{\bm{k}}^\dagger d_{-\bm{k}}^\dagger + {\rm H.c.}\bigr)  \nonumber\\
&\ -\cos k_y \bigl(a_{\bm{k}}^\dagger c_{\bm{k}}
+b_{\bm{k}}^\dagger d_{\bm{k}}
+a_{\bm{k}}^\dagger c_{-\bm{k}}^\dagger
+b_{\bm{k}}^\dagger d_{-\bm{k}}^\dagger + {\rm H.c.}\bigr) , \\
 h_{\bm{k}}^x&=\cos (k_x+k_y) \bigl(a_{\bm{k}}^\dagger d_{\bm{k}}
+a_{\bm{k}}^\dagger d_{-\bm{k}}^\dagger
+ {\rm H.c.}\bigr)  \nonumber\\
&\ \ \ \ +\cos (k_x-k_y) \bigl(b_{\bm{k}}^\dagger c_{\bm{k}}
+b_{\bm{k}}^\dagger c_{-\bm{k}}^\dagger + {\rm H.c.}\bigr).
\end{align}
The first term in Eq.~(\ref{eq:4}) is the zero-th order energy, denoted by $E_0$, which is independent of the angle $\varphi$ at $J=1$, as mentioned previously. By applying the Bogoliubov transformation, we obtain a diagonalized form of the Hamiltonian as
\begin{align}
 H_{\rm SW}&=E_0+\Delta E +\sum_{\bm{k},\eta\eta'=\pm}
\omega_{\bm{k}}^{\eta\eta'}\alpha_{\bm{k}}^{\eta\eta'\dagger}\alpha_{\bm{k}}^{\eta\eta'},
\label{eq:sw}
\end{align}
where $\alpha_{\bm{k}}^{\eta\eta'}$ is the boson operator and subscripts $\eta$ and $\eta'$ take $\pm$. The energy dispersions are given as
\begin{align}
 \omega_{\bm{k}}^{\eta\eta'}/(2J_z)&=2S\sqrt{X_{\bm{k}}^\eta+\eta' Y_{\bm{k}}^\eta}
\end{align}
with
\begin{align}
X_{\bm{k}}^\eta=&
\left \{ J+1-(J-1)\cos 2\varphi \right  \}^2 
-2\eta J\cos k_x\cos k_y\cos^2\varphi \nonumber \\ 
& \times \left \{ J+1-(J-1)\cos 2\varphi \right \}, 
\end{align}
and 
\begin{align}
Y_{\bm{k}}^\eta=&2 \left ( \cos^2\varphi+J\sin^2\varphi \right )
\nonumber\\
&\!\!\!\!\!\!\!\!\!\!\!\!\!\!\!\!\!\!
\times\sqrt{4J^2\sin^2 k_x\sin^2 k_y\cos^4\varphi+(\cos k_x+\eta \cos k_y)^2\sin^4\varphi}.
\end{align}
In the cases of $\varphi=0$ and $\pi/2$, the dispersion relations are reduced to
\begin{align}
\omega_{\bm{k}}^{\eta\eta'}/(2J_z)=2S\sqrt{1+\eta\eta' J\cos(k_x+\eta k_y)} , \label{eq:9}
\end{align}
and 
\begin{align}
\omega_{\bm{k}}^{\eta\eta'}/(2J_z)=2S\sqrt{J \left \{ J +\eta' (\cos k_x +\eta \cos k_y)/2 \right \}} , 
\end{align}
respectively.
The second term in Eq.~(\ref{eq:sw}) is a correction due to the zero-point vibration given by
\begin{align}
 \Delta E&=\frac{N}{8}\int_{-\pi}^\pi\int_{-\pi}^\pi\frac{dk_x dk_y}{(2\pi)^2}
\sum_{\eta\eta'=\pm}\omega_{\bm{k}}^{\eta\eta'}\nonumber\\
&\ \ -2J_zNS(\cos^2\varphi+J\sin^2\varphi).
\end{align}

\begin{figure}[!t]
\begin{center}
\includegraphics[scale=0.55]{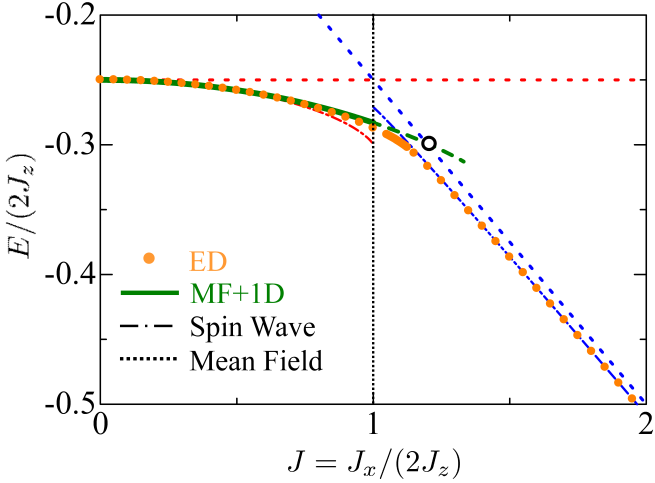}
\caption{
Ground state energy at $T=0$ obtained by using several methods. Dotted and dashed-dotted lines represent results obtained by the MF approximation and the spin-wave method, respectively. Solid line represents the result by MD+1D method.
Green broken line shows the coexistent area in the first-order phase transition. Filled circles represent the results by the Lanczos method in a cluster of $N=32$.}
\label{check_ED}
\end{center}
\end{figure}
In Fig.~\ref{check_ED}, the ground state energy including the quantum correction, $E_0+\Delta E$, is plotted as a function of $J$. Reductions from the MF energies are remarkable around $J=1$. At $J=1$, energy for the limit of $J \rightarrow 1-0$ is lower than that for $J \rightarrow 1+0$. Energies between the two configurations are shown in Fig.~\ref{check_sw_phi} as a function of the rotation angle $\varphi$. Energy reduction due to the zero-point vibration is the largest at $\varphi=0$. That is, the $\langle T^z_{(\pi, \pi)} \rangle$ order is stabilized among the continuous degenerate configurations. This is attributed to the one-dimensional character of the spin-wave dispersion relation which gives rise to the large excitation density at low energies.

\subsection{Jordan-Wigner Method with MF Approximation}
In order to analyze the orbital state beyond the spin-wave approximation, we use a combined method of the Jordan-Wigner transformation and the MF approximation. We term this method MF+1D for simplicity. The MF approximation is applied to the NN interaction, $T_i^z T_j^z$, and the one-dimensional chain under the MF is analyzed by the Jordan-Wigner transformation.~\cite{chen07} This treatment is justified for $J\ll 1$.

By introducing the MF approximation in the NN interaction, the Hamiltonian for each diagonal chain is mapped onto the independent one-dimensional transverse Ising model given by 
\begin{align}
 H_{\rm MF-1D}/(2J_z)&=J\sum_{<ij>'}T_i^x T_j^x -h\sum_i T_i^z+L\mean{T^z}^2 ,
\label{eq:33}
\end{align}
where the Neel-type MF, $\means{T^z} \equiv \means{T^z_{(\pi,\pi)}}$, is assumed. A symbol $<ij>'$ represents a NN $ij$ pair in a one-dimensional chain, and $L$ is a size of a chain.
The transverse field is given by 
\begin{align}
 h=-2\mean{T^z}.\label{eq:5}
\end{align}
By introducing the Jordan-Wigner transformation, the PS operators are represented by a spin-less fermion operator $c_i$ such as $T_i^z=c_i^\dagger c_i-\frac{1}{2}$ and others. The fermion model is diagonalized by using the Bogoliubov transformation, and the following Hamiltonian is obtained;
\begin{align}
 H_{\rm MF-1D}/(2J_z)&=\sum_{k}E_{k}\left( \alpha_{k}^\dagger \alpha_{k}-\frac{1}{2} \right)+L\mean{T^z}^2,
\end{align}
where $\alpha_i$ is a fermion operator introduced by the Bogoliubov transformation defined by 
\begin{align}
 \begin{cases}
  \alpha_k=u_k c_k-v_k c_{-k}^\dagger , \\
  \alpha_{-k}^\dagger=u_{k} c_{-k}^\dagger+v_{k}^* c_{k} , 
 \end{cases}
\end{align}
with the coefficients given by 
\begin{align}
 \begin{cases}\displaystyle
  u_k^2=\frac{1}{2}\left( 1-\frac{h-\frac{J}{2}\cos k}{E_k} \right) , \\\displaystyle
  |v_k|^2=\frac{1}{2}\left( 1+\frac{h-\frac{J}{2}\cos k}{E_k} \right).
 \end{cases}
\end{align}
The eigen energy for the fermion is given as
\begin{align}
 E_{k}=\sqrt{\frac{J^2}{4} -Jh\cos k +h^2},\label{eq:37}
\end{align}
where $k$ is the wave vector along the $\langle 11 \rangle$ or $\langle {\bar 1} 1\rangle$ directions. 
We calculate the expectation value $\mean{T^z}$  by solving the self-consistent equations in Eq.~(\ref{eq:33}) and Eq.~(\ref{eq:5}). At zero temperature, we have 
\begin{align}
 \means{T^z}_0&=\frac{1}{L}\sum_i\means{c_i^\dagger
 c_i}_0-\frac{1}{2}
 \nonumber\\
&=\frac{1}{2\pi}\int_0^\pi  \frac{-h+\frac{J}{2}\cos k}{\sqrt{\frac{J^2}{4}-Jh
 \cos k+h^2}}dk,\label{eq:36}
\end{align}
where $\mean{\cdots}_0$ represents an expectation value at zero temperature. The self-consistent equation is given as 
\begin{align}
 h&=-2\means{T^z}_0=-\frac{1}{\pi}\int_0^\pi  \frac{-h+\frac{J}{2}\cos k}{\sqrt{\frac{J^2}{4}-Jh
 \cos k+h^2}}dk. \label{eq:34}
\end{align}
At finite temperature, the partition function and the free energy are obtained from Eq.~(\ref{eq:33}) as 
\begin{align}
 Z&=e^{-(2J_z)L\beta \mean{T^z}^2}\prod_k 2\cosh\left[ (2J_z)
 \frac{\beta E_k}{2} \right]
\end{align}
and 
\begin{align}
 F/(2J_z L)&=\mean{T^z}^2 -\frac{T}{2J_z}\frac{1}{L}\sum_k  \ln \left[
2\cosh\left(\frac{2J_z}{T} 
 \frac{E_k}{2} \right)
 \right], 
 \label{eq:35}
\end{align}
respectively. 

\begin{figure}[t]
\begin{center}
\includegraphics[scale=0.5]{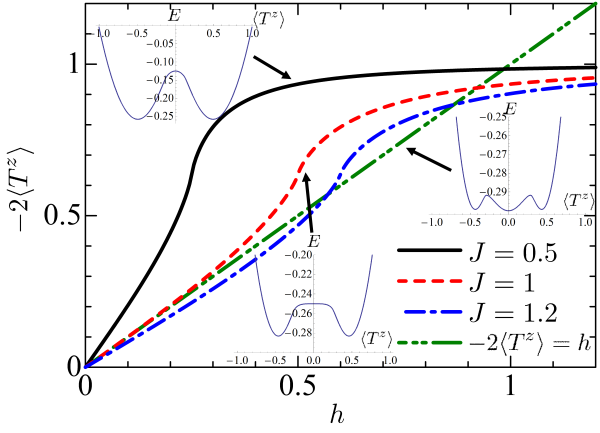} 
\caption{
The left and right hand-sides of the self-consistent equation in Eq.~(\ref{eq:34}) at $T=0$. Energy versus $\langle T^z \rangle$ curves for each $J$ are also shown.}
\label{check_mf1d_hs}
\end{center}
\end{figure}
In Fig.~\ref{check_mf1d_hs}, we plot the left- and right-hand sides of the self-consistent equation in Eq.~(\ref{eq:34}) in a region of the positive effective field ($h \ge 0$). When $J$ is less than one, the equations have solutions at $h=0$ and at one positive value of $h$. At $J=1$, slopes in the two curves coincide with each other at $h=0$. Additional solution appears in the case of $1<J\lesssim 1.35 $.
The energies are calculated as functions of $\langle T^z \rangle$ (see the insets of Fig.~\ref{check_mf1d_hs}). The energy has three minima in the region of $1 \le J\le 1.35$. In particular, in $1.2< J<1.35$, $E$ shows absolute minima at $\langle T^z \rangle=0$. A discontinuous change in the stable $\langle T^z \rangle $ at $J=1.2$ implies the first-order phase transition. 
This is attributed to an existence of an inflection point in the $h$-$\mean{T^z}$ curve, i.e. the magnetization curve in the transverse Ising model at $T=0$. It is well known that this model shows a second-order phase transition at $T=0$, a quantum critical point, at a certain value of $h/J$, where the magnetic susceptibility diverges due to an inflection point in the magnetization curve. In this sense, the present first-order phase transition originates from competition between the directional dependent PS interactions, that is, $T^z_i T^z_{j}$ along the NN bonds and $T^x_iT^x_{j}$ along the NNN bonds in the compass model.

Results at $T=0$ obtained by the 1D+MF method together with the results by other methods are summarized in Fig.~\ref{check_ED}. Broken line represents a region where solutions of $\mean{T^z}=0$ and $\mean{T^z} \ne 0$ coexist, and an open circle indicates a point where the absolute minima change from $\mean{T^z} \ne 0$ to $\mean{T^z}=0$. At $J=1$, where the continuous degeneracy exists in the MF solutions, the $\mean{T^z}$ order is realized in the 1D+MF method. These results are consistent with the results in the spin-wave approximation. 

\begin{figure}[t]
\begin{center}
\includegraphics[scale=0.4]{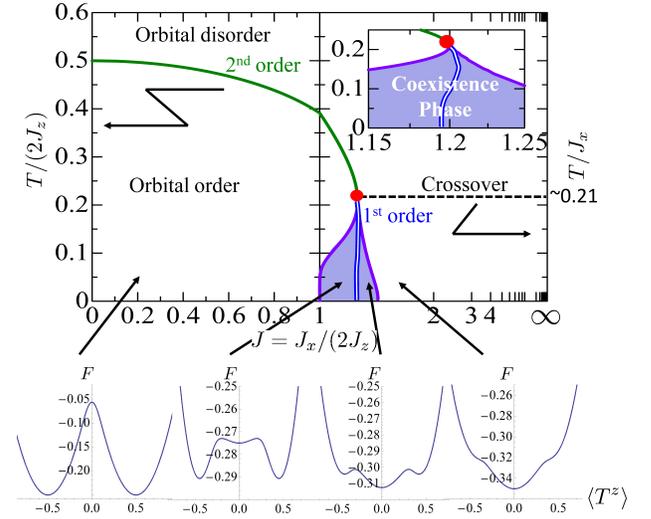}
\caption{
Phase diagram obtained by using the MF+1D method. A red circle represents the tricritical point. Broken line indicates the temperature below which the one-dimensional $T^x$ correlation developed. The lines in $J<1$ and $J>1$ are plotted in the different scales at left and right figure, respectively. A shaded area implies the coexistence region in the first-order phase transition. Energy versus $\langle T^z \rangle$ curves at $(J,T/(2J_z))=$(0.2,0.05), (1.1,0.05), (1.25,0.05) and (1.4,0.05) are also shown. The inset shows the extension around $J=1.2$ and $T/J_x=0.1$.
}
\label{check_mf1d_phase}
\end{center}
\end{figure}
Next we present the phase diagram at finite temperature. We suppose an existence of the tricritical point at a certain ($J, T$), since the second-order phase transition appears at $J=0$ where the model is reduced to the two-dimensional Ising model, and the first-order phase transition is confirmed at $T=0$ as explained above. We expand the free energy in Eq.~(\ref{eq:35}) at the vicinity of $\mean{T^z}=0$ as 
\begin{align}
 F/(2J_z L)&=f_0 + \mean{T^z}^2 f_2 +\mean{T^z}^4 f_4+\cdots,
\end{align}
with coefficients given by
\begin{align}
 f_0&=-t \ln\left[ 2 \cosh\left( \frac{J}{2 t} \right) \right] , \\
f_2&=1-\frac{1}{4Jt}{\rm sech}\left( \frac{J}{4t} \right)\left[ J+2t\sinh\left( \frac{J}{2t} \right) \right] , \\
f_4&=\frac{3}{32t^3}{\rm sech}^4\left( \frac{J}{4t} \right)
-\frac{1}{J^3}\tanh\left( \frac{J}{4t} \right)  \nonumber\\
&\ \ \ \ \ -\frac{1}{16J^2 t^3}{\rm sech}^2\left( \frac{J}{4t} \right)\left[ J^2-4t^2-4Jt\tanh\left( \frac{J}{4t} \right) \right] , 
\end{align}
where $t=T/(2J_z)$. The second-order phase-transition point is given by $f_2=0$ and $f_4 > 0$, and the tricritical point is given by $f_2=f_4=0$. The finite-$T$ phase diagram is presented in Fig.~\ref{check_mf1d_phase}. We also plot the first-order phase transition points and the hysteresis region determined by the free energy in Eq.~(\ref{eq:35}). Broken line represents a crossover, below which the one-dimensional $T^x$ correlation is developed, and is numerically determined by a peak in the specific heat $C=-T(\partial^2 F)/(\partial T^2)$ at $\means{T^z}=0$. This line does not depend on $J$, because the present model in Eq.~(\ref{eq:33}) is reduced to the independent one-dimensional Ising model, when $\langle T^z \rangle=0$.
Both the results obtained by the MF+1D method and the results by the classical MC method (see Fig.~\ref{check_classical_phase}) show that the phase transition for $\means{T^z_{(\pi,\pi)}}$ is changed to be the second order to the first order through the tricritical point with increasing $J$. One discrepancy is seen at $J \sim 1$ in low temperatures; in the quantum phase diagram, the $\means{T^z_{(\pi,\pi)}}$ order is realized up to $J \sim 1.2$ even at $T=0$. This is a kind of order by fluctuation phenomena due to the quantum fluctuation. 

\subsection{Exact Diagonalization Method}

\begin{figure}[!t]
\begin{center}
\includegraphics[scale=0.5]{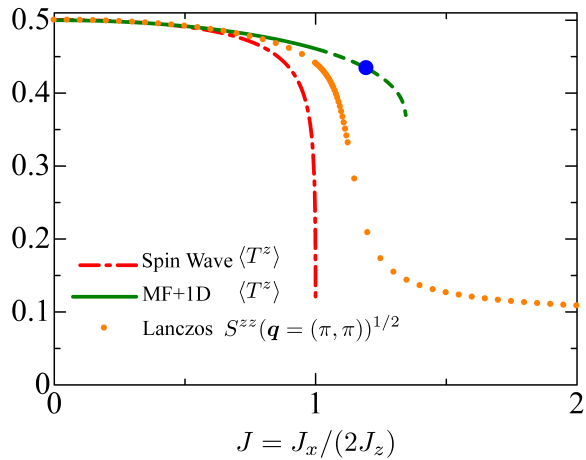}
\caption{
Amplitude of the $\means{T^z_{(\pi,\pi)}}$ order obtained by the MF+1D method (bold line) and that by the spin-wave approximation (dashed-dotted line). Broken line represents the results for the coexistent region in the first-order phase transition obtained by the MF+1D method. The square root of the orbital correlation function $S^{zz}(\pi, \pi)$ calculated in the Lanczos method are plotted by filled circles.
}
\label{check_moment}
\end{center}
\end{figure}
To examine the orbital state at $T=0$ in more detail, we adopt the exact diagonalization method based on the Lanczos algorithm. We use a $4\sqrt{2}\times 4\sqrt{2}$-site cluster, where edges are parallel to the $\langle 11\rangle$ and $\langle 1\bar{1}\rangle$ directions, with the periodic boundary condition. Calculated energy is plotted as a function of $J$ in Fig.~\ref{check_ED}. The results are good agreement with the results  obtained by the MF+1D method for $J \lesssim 1$, and with the results by the spin-wave approximation except for a region of $J \sim 1$.

The square root of the orbital correlation function $S^{zz}(\pi, \pi)$ calculated by the Lanczos method is compared with the ordered moment of the $\langle T^z_{(\pi, \pi)} \rangle$ order obtained by other methods (see Fig.~\ref{check_moment}). The results obtained by three methods coincide with each other in a region of small $J$. However, at the vicinity of $J=1$, large discrepancies between the three results are observed.
Obtained ordered moment in the MF+1D method is larger than that in the spin-wave approximation. This tendency might be due to underestimation (overestimation) for the fluctuation in the MF+1D method (spin-wave approximation). Data for the correlation function obtained by the Lanczos method are located between the results of $\langle T^z \rangle$ by the MF-1D method and the spin wave approximation.

\subsection{Quantum Monte Carlo Simulation}
\begin{figure}[!t]
\begin{center}
\includegraphics[scale=0.48]{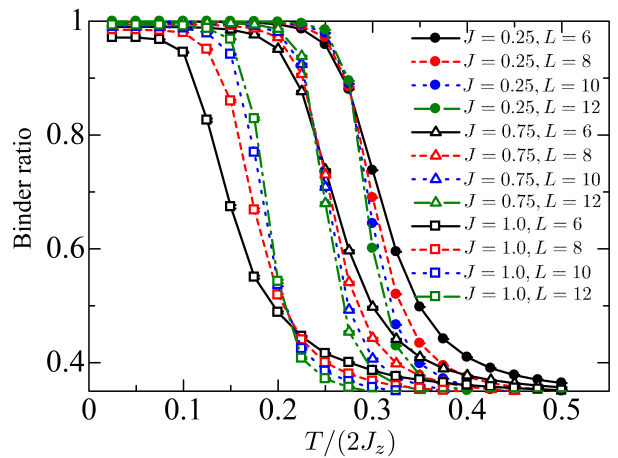}
\caption{
Temperature dependence of the binder ratio (see Eq.~(\ref{eq:48})) obtained by the QMC method with $L= 6-12$ where $5\times 10^5$ MC steps are used for measurements. Statistical errors are estimated from 64 independent runs.
}
\label{check_BIN}
\end{center}
\end{figure}
\begin{figure}[!t]
\begin{center}
\includegraphics[scale=0.45]{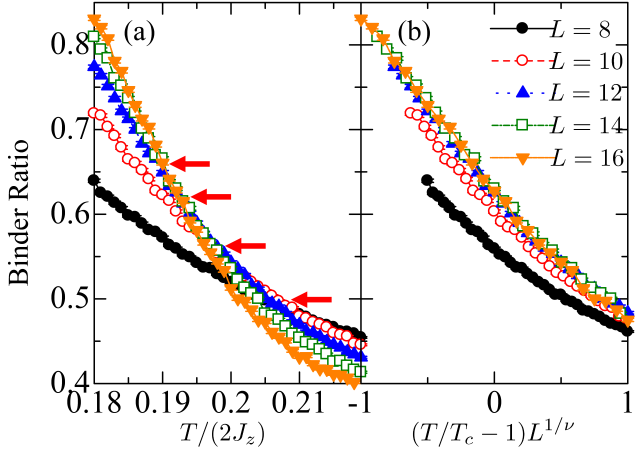}
\caption{
(a) Detailed temperature dependence of the Binder ratio at $J=1.0$ with $L=8-16$. The arrows indicate the crossing points for the two curves in the $L$ and $L+2$ site clusters. (b) Scaling plot of the Binder ratio. We chose $\nu=1$ and $T_c/(2J_z)=0.192$.
}
\label{check_BIN2}
\end{center}
\end{figure}
We present the numerical results obtained by the quantum Monte Carlo (QMC) method. Before showing the numerical results, we touch signs of the exchange constants in the Hamiltonian in relation to the negative sign problem. Signs of the two exchange constants are positive  in a view point of the perturbational calculation, but these signs can be reversed by the following way. A checkerboard lattice is decomposed into the NN and NNN bond networks where $T^z$ and $T^x$ components are only concerned, respectively. Since the two networks are bipartite, signs of the exchange interactions can be reversed by introducing the unitary transformations of $T_i^z\rightarrow U_y^{-1}(\pi) T^z_i U_y(\pi)$ for the sites $(i_x, i_y)$ of $i_x+i_y$=odd, and $T_i^x\rightarrow U_y^{-1}(\pi) T^x_i U_y(\pi)$ for the sites of $i_x$=odd. The unitary matrix $U_y(\pi)$ represents the $\pi$ rotation around the $T^y$ axis. In the simulations, we introduce the above transformation, and the negative sign problem does not appear. In the manuscript, we choose signs of the exchange constants to be positive.

We perform the continuous imaginary-time method with the loop algorithm in the ALPS library.\cite{ALPS,Todo} We use $\sqrt{2}L\times \sqrt{2}L$-site clusters ($L=6-16$), where edges are parallel to the $\langle 11\rangle$ and $\langle 1\bar{1}\rangle$ directions, with the periodic boundary condition. To calculate the physical quantities, $6\times 10^5-5\times 10^8$ MC steps are used. Statistical errors are estimated from 4-64 independent runs.

In the region of $J \lesssim 1$, increasing of the correlation function $S^{zz}(\pi, \pi)$ at a certain temperature is observed (not shown in figure).  The results indicate a possibility of the $\means{T^z_{(\pi,\pi)}}$ order. In order to determine the critical point of this order, we utilize the Binder ratio defined by
\begin{align}
 g&=\frac{\means{T^{z2}_{(\pi,\pi)}}^2}{\means{T^{z4}_{(\pi,\pi)}}}\label{eq:48}.
\end{align}
In principle, this quantity does not depend on the cluster size at critical temperature. This is shown by utilizing the scaling relation given by 
\begin{align}
 g=f_g \left [ L^{1/\nu}(T-T_c) \right ],
\end{align}
where $\nu$ is the critical exponent for the correlation length and $f_g$ is the scaling function. Figure~\ref{check_BIN} presents the temperature dependence of $g$ for the several values of $J$ and $N$. In the data sets for $J=0.25$ and $0.75$, the crossing points are observed. Deceasing of $T$ at the crossing point with increasing $J$ is consistent with the results obtained by the classical MC method and the MF+1D method (see Figs.~\ref{check_classical_phase} and~\ref{check_mf1d_phase}).

We focus on values of $g$ at the crossing point, termed $\tilde{g}_c$, in Fig.~\ref{check_BIN}. It is known that, in general, $\tilde{g}_c$ does not depend on $J$, and about 0.85 for the two dimensional Ising universality class.\cite{Schmid} This figure shows that $\tilde{g}_c$'s at $J=0.25$ and 0.75 are close to this value, and the transitions are expected to belong to the two-dimensional Ising universality class. As for the case at $J=1$, detailed results of $g$ are presented in Fig.~\ref{check_BIN2}(a). Up to the results of $L=16$, curves for different $L$ do not cross with each other at same point. A value of $g$ at the crossing point in the $L$ and $L+2$ site clusters increases with increasing $L$, and might approach to 0.85 in the case of larger $L$. On the other hand, as shown in Fig.~\ref{check_BIN2}(b), the Binder ratios plotted as functions of $(T/T_c-1)L^{1/\nu}$ are fitted by a single curve in the case of $L\geq 12$, and $\nu=1$ and $T_c/(2J_z)=0.192$ are obtained. The obtained value of $\nu$ is consistent with the two dimensional Ising universality. From these analyses, we suppose that, at $J=1$, the second order phase transition of $\means{T^z_{(\pi,\pi)}}$ is realized, and larger size clusters are required to examine $g_c$ than the clusters where the finite-size scaling for $\nu$ works well.

\begin{figure}[!t]
\begin{center}
\includegraphics[scale=0.5]{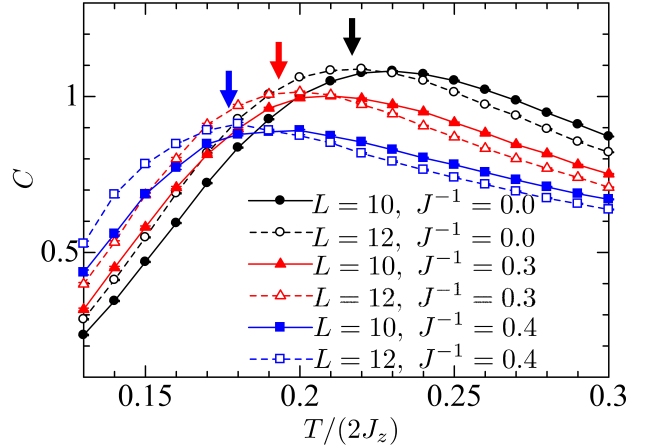}
\caption{
Temperature dependence of the specific heat obtained by the QMC method, where $4\times 10^6$ MC steps are used for measurements.
}
\label{check_Cv_Jx}
\end{center}
\end{figure}
In the region of $ J \gtrsim 1$, on the other side, the calculated correlation function $S^{zz}(\pi,\pi)$ does not show remarkable development with decreasing $T$. We examine the crossover temperature below which the one-dimensional $T^x$ correlation develops for $J \gtrsim 1$, as suggested in other calculation methods. The temperature dependences of the specific heat for several $J$ and $N$ are shown in Fig.~\ref{check_Cv_Jx}. The maxima of the specific heats are indicated by small arrows. The crossover temperature where $C$ takes its maximum decreases with decreasing $J$. This tendency is similar to that observed by the classical MC method (see Fig.~\ref{check_classical_phase}), but is in contrast to that by the MF+1D method (see Fig.~\ref{check_mf1d_phase}) where the crossover temperature does not depended on $J$. This is attributed to the approximation in the MF+1D method in which the each one-dimensional chain is treated to be independent in this region.

\begin{figure}[!t]
\begin{center}
\includegraphics[scale=0.45]{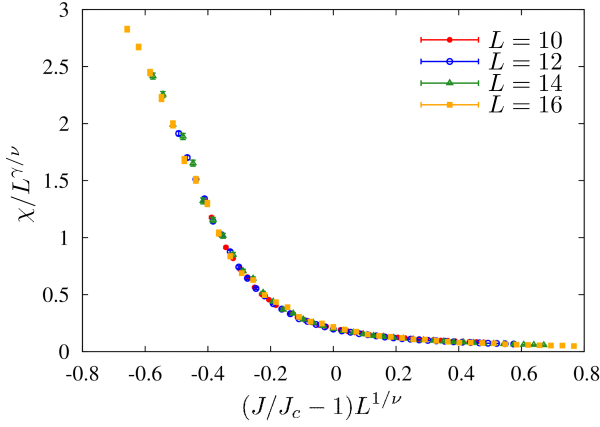}
\caption{
A scaling plot for the susceptibility at $T/(2J_z)=0.15$. We chose $\gamma=7/4$, $\nu=1$ and $J_c/(2J_z)=1.095$.
}
\label{0.3}
\end{center}
\end{figure}
\begin{figure}[!t]
\begin{center}
\includegraphics[scale=0.45]{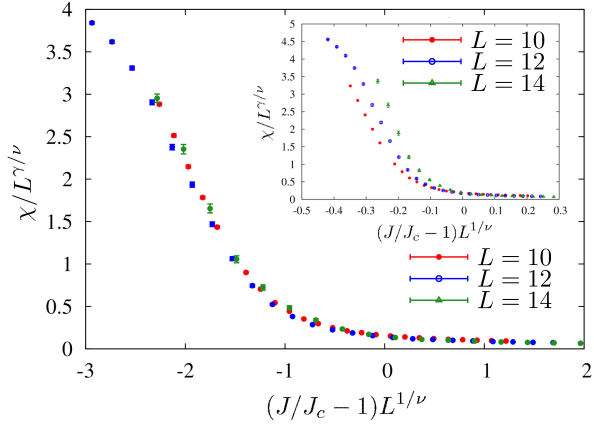}
\caption{
A scaling plot for the susceptibility at $T/(2J_z)=0.125$. We chose $\gamma=1$, $\nu=5/9$ and $J_c/(2J_z)=1.089$. Inset shows a scaling plot where $\gamma=7/4$, $\nu=1$ and $J_c/(2J_z)=1.088$ are chosen.
}
\label{0.25}
\end{center}
\end{figure}
\begin{figure}[!t]
\begin{center}
\includegraphics[scale=0.45]{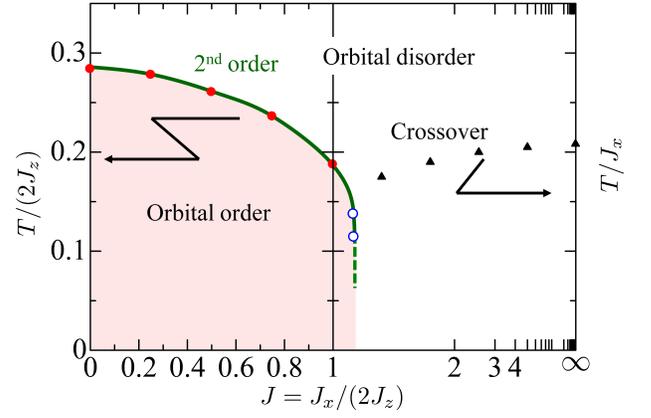}
\caption{
Phase diagram obtained by the QMC method. The lines in $J<1$ and $J>1$ are plotted in the different scales at left and right figure, respectively. Filled and open circles for the second-order phase transition are obtained by the finite size scalings of the Binder ratio and the susceptibility, respectively. Triangles for the crossover points are determined by the specific heat. 
}
\label{check_QMC_phase}
\end{center}
\end{figure}
Next, we introduce careful examinations for the orbital states around $J=1$. An accuracy of the MC simulation is checked by calculating the auto-correlation time for the auto-correlation function defined by 
\begin{align}
C^{zz}(\tau)=\means{T_{(\pi,\pi)}^z(\tau) T_{(\pi,\pi)}^z(0)}-\means{T_{(\pi,\pi)}^z(\tau)} \means{T_{(\pi,\pi)}^z(0)}, 
\end{align}
where $T_{(\pi,\pi)}^z(\tau)$ is the staggered orbital moment for the $\tau$-th configuration in the Markov chain.\cite{Evertz} In the simulations where the MC steps are taken to be $10^9$, saturations for the auto-correlation times are observed above $T/(2J_z)=0.125$, but not observed below $T/(2J_z)=0.1$, and the numerical results obtained above $T/(2J_z)=0.125$ are reliable. 
The orbital susceptibility at ${\bm q}=(\pi,\pi)$ defined by 
\begin{align}
 \chi&=\int_0^\beta d\tau [\means{e^{\tau H}T_{(\pi,\pi)}^z e^{-\tau H} T_{(\pi,\pi)}^z}-\means{T_{(\pi,\pi)}^z}^2],
\end{align}
is calculated at $T/(2J_z)=0.125$ and $0.15$. The MC steps are chosen to be $1\times 10^8$ for $T/(2J_z)=0.15$ and $5\times 10^8$ for $T/(2J_z)=0.125$, and averaged values in the 4-times measurements are calculated.
When temperature and system size are fixed, the susceptibility calculated as a function of $J$ shows abrupt increase at a certain value of $J$ which is termed $J_c(T, L)$. When we assume that this points are the continuous critical points in the $J-T$ plane, we expect a linear correspondence between $J$ and $T$ near the points. Therefore, from the conventional scaling form for the susceptibility as a function of $T$, given by
\begin{align}
 \chi=L^{\gamma/\nu}f_{\chi}[L^{1/\nu}(T-T_c)],
\end{align}
where $\gamma$ is the critical exponent and $f_{\chi}$ is the scaling function, the following scaling relation as a function of $J$ is expected 
\begin{align}
 \chi=L^{\gamma/\nu}f^J_{\chi}[L^{1/\nu}(J-J_c)],
\end{align}
where we introduce a scaling function $f^J_\chi$. We suppose that, near the critical points, $\chi/L^{\gamma/\nu}$ versus $L^{1/\nu}(J-J_c)$ data for several $L$ are on a single curve.

A scaling plot at $T/(2J_z)=0.15$ is shown in Fig.~\ref{0.3} where $(\gamma, \nu)=(7/4, 1)$, expected from the two-dimensional Ising universality class, and $J_c=1.095$ are used. The optimized values obtained by the least-squares fit are $(\gamma, \nu)=(1.5\pm 0.8, 1.07\pm 0.11)$, and $J_c=1.092\pm 0.004$. Scaling plot works well; numerical data obtained by several $N$ are fitted by a scaling function. This analysis indicates that the second-order phase transition line continues from $(T/(2J_z),J)=(0.28,0)$ to $(0.15,1.092)$.

On the contrary, the scaling analyses with the exponents $(\gamma, \nu)=(7/4, 1)$ do not fit the numerical data at $T/(2J_z)=0.125$, as shown in the inset of Fig.~\ref{0.25}.
A different plot, where $(\gamma, \nu)=(1, 5/9)$ and $J_c=1.089$ are used, is presented in Fig.~\ref{0.25} for the data at $T/(2J_z)=0.125$. 
The optimized values by the least-squares fit are $(\gamma, \nu)=(0.8\pm 0.5, 0.58\pm 0.08)$ and $J_c=1.094\pm 0.006$. 
All data obtained in different $N$ are almost fitted by a single function. 
The values $(\gamma, \nu)=(1, 5/9)$ are the critical exponents for the two-dimensional tricritical Ising universality class obtained by the $c=7/10$ conformal field theory.~\cite{Friedan,Landau81}

Phase diagram obtained by the finite-size scaling analyses in the Binder ratio and the susceptibility is given in Fig.~\ref{check_QMC_phase}. As explained above, through the scaling analyses, we propose a possibility that the tricritical point exists around $(T/(2J_z),J)=(0.125,1.094)$. Because of an accuracy of the QMC simulation, the first-order phase transition expected below $(T/(2J_z),J)=(0.125,1.094)$ is not confirmed by the numerical simulation. However, an existence of the tricritical point is reasonable by taking into account of the results obtained by other methods of the classical MC simulation and the MF+1D method shown in Figs.~\ref{check_classical_phase} and \ref{check_mf1d_phase}.

\begin{figure}[!t]
\begin{center}
\includegraphics[scale=0.59]{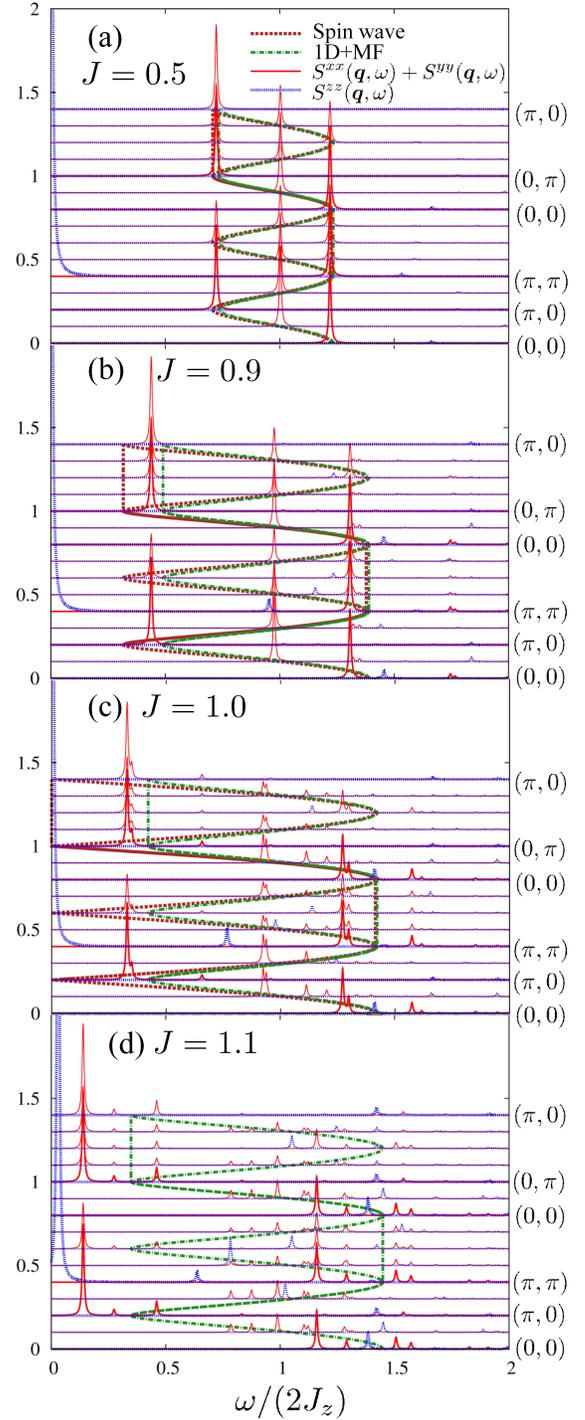}
\caption{
Spin wave dispersion relations obtained by the spin-wave approximation (dashed lines), and by the MF+1D method (dashed-dotted lines). Three dimensional plots for the dynamical PS correlation functions for the transverse component  $S^{xx}(\bm{q},\omega)+S^{yy}(\bm{q},\omega)$ (bold lines) and those for the longitudinal component $S^{zz}(\bm{q},\omega)$ (dotted lines) obtained by the Lanczos method are also shown. Parameters are chosen to be (a)$J=0.5$, (b)$J=0.9$, (c)$J=1.0$ and (d)$J=1.1$.
}
\label{spectra}
\end{center}
\end{figure}
\begin{figure}[!t]
\begin{center}
\includegraphics[scale=0.6]{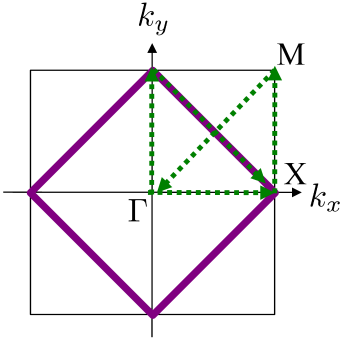}
\caption{The first Brillouin zone for the checkerboard lattice. Momenta in Fig.~\ref{spectra} are varied along the arrows. Bold lines represents $k_x \pm k_z=\pm \pi$ (see text).
}
\label{check_dep}
\end{center}
\end{figure}

\section{Dynamical Orbital State}
In this section, we present numerical results for the excitation spectra in the checkerboard compass model. The excitation spectra are calculated by using the spin-wave approximation, the MF+1D method and the continued fraction expansion method based on the Lanczos method. In the  spin-wave approximation and the MF+1D method, the $\langle T^z_{(\pi, \pi)} \rangle$ order is assumed. Results are presented in the Brillouin zone for the orbital disordered phase. 

The excitation spectra obtained by the spin-wave approximation and the MD+1D method are explicitly given by 
\begin{align}
 \omega_{{\rm SW};\bm{k}}^{(\pm)}/(2J_z)&=\sqrt{1- J\cos(k_x\pm k_y)},\label{eq:38}
\end{align}
and
\begin{align}
 \omega_{{\rm MF};\bm{k}}^{(\pm)}/(2J_z)=\sqrt{\frac{J^2}{4} -Jh\cos
 (k_x\pm k_y) +h^2},\label{eq:8}
\end{align}
from Eq.~(\ref{eq:9}) and Eq.~(\ref{eq:37}) respectively. In the limit of $J\ll 1$, $\omega_{{\rm SW};\bm{k}}^{(\pm)}$ coincides with $\omega_{{\rm MF};\bm{k}}^{(\pm)}$ under the assumption of $h=-2\mean{T^z}= \pm 1$. This is reasonable because the two approximations are equivalent with each other in this limit. Spin waves show one-dimensional character; dispersions appear along $\langle 11 \rangle$ or $\langle 1{\bar 1} \rangle$ directions in the Brillouin zone. This is because, PS fluctuations in the $\langle T^z_{(\pi, \pi)} \rangle$ ordered state are caused by the interactions between $T^x$ along the diagonal directions on the checkerboard lattice. 
In the Lanczos method, we calculate the dynamical correlation function given by
\begin{align}
 S^{ll}(\bm{q},\omega)&=-\frac{1}{\pi}{\rm Im}\means{T^l_{\bm{q}} \frac{1}{\omega-H+E_g+i\eta}T^l_{\bm{-q}}}
\end{align}
where $l=(x,y,z)$, $E_g$ is the ground-state energy and $\eta$ is an infinitesimal constant. In the numerical calculations, the system size is taken to be $N=32$ and $\eta$ is chosen to be $\eta/(2J_z)=0.01$.

We show the excitation spectra for several $J$ obtained by the three methods in Fig.~\ref{spectra}, where the momenta are varied along arrows shown in Fig.~\ref{check_dep}. Three results show good agreement with each other in the case of $J=0.5$. Discrepancies between the three results are remarkable around $J=1$. In particular, noticeable differences are observed in the lowest energy excitations; the lowest excitation energy by the spin wave approximation (the 1D+MF method) is the lowest (highest) among the three results. A zero-energy peak in $S^{zz}(\bm{q})$ at $\bm{q}=(\pi,\pi)$ is due to the static staggered correlation for $T^z$ in the ground state. At $J=1$, $\omega_{{\rm SW};\bm{k}}^{(\pm)}$ shows gapless excitations. This is not the Goldstone mode but is due to the linear spin wave approximation, and reflect the continuous degeneracy in the MF solutions at $J=1$. We expect the dispersions are gapful when the higher order corrections in the spin wave approximation are taken into account. 

Let us focus on the results obtained by the Lanczos method. The present results in $J=0.5$ well reproduce the results obtained by other two methods. With increasing $J$ up to around $J=1$, except for the lowest peaks, almost all peak intensities are diminished and a number of small incoherent peaks appear. This might be attributed to the magnon-mangnon interaction which becomes remarkable when the system approaches to $J=1$. This result is related to the amplitude of the $\langle T^z_{(\pi, \pi)} \rangle$ order as well as the corresponding correlation function shown in Fig.~\ref{check_moment}, where their reductions are due to the spin wave excitations. 
As for the lowest coherent peaks, even in the results by the Lanczos method beyond the spin wave approximation, their energies are almost flat along the lines of $k_x \pm k_y= \pm \pi$. These are shown in Fig.~\ref{check_dep} and correspond to the momenta for the PS configurations stabilized in $J \gtrsim 1.35$.  In this sense, the softening of the lowest coherent peaks implies a precursor of this PS configuration, although the phase transition  at $T=0$ is of the first order.

\section{Discussion and Summary}

We discuss the present results in the checkerboard orbital compass model in comparison with the square lattice orbital compass model (SLCM). There are a number of theoretical studies in the orbital compass model on a square lattice defined by 
\begin{align}
 H=J_z\sum_{<ij>_z}T_i^z T_j^z+J_x\sum_{<ij>_x}T_i^x T_j^x ,
\end{align}
where the first and second terms are the NN interactions along the horizontal and vertical directions on a square lattice, respectively. There are the generators, which are similar to Eq.~(\ref{eq:pl}) in the present model, defined as $P_l=\prod_{i\in l} T_i^x$ and $Q_m=\prod_{i\in m} T_i^z$ , where $l$ and $m$ indicate the $l$-th row and the $m$-th column on a square lattice, respectively. The Hamiltonian commutes with $P_l$ and $Q_m$ for any $l$ and $m$. The ground and excited states at $T=0$ have been studied by several methods. It was shown in the anisotropic case, i.e. $J_x \ne J_z$ that on a $L \times L$-site lattice, the low energy spectrum consisting of $2^L$ states collapse exponentially fast with each other with increasing a system size. The first-order phase transition might occur at the symmetric point of $J_x=J_z$.\cite{chen07} In contrast to SLCM, in the present checkerboard model, the two states realized in the large and small limits of $J=J_x/(2J_z)$ are not symmetrical with each other. The N\'eel-type symmetry-broken state is stabilized in the region of $J \lesssim 1$, and the $2^{2L}$-fold degenerate staggered $T^x$ ordered state along the diagonal directions appear from $J=\infty$ down to around $J=1$. The first-order transition between the two occurs around $J=1.35$ at $T=0$. Just at $J=1$, accidental continuous degeneracy is observed in the classical ground state in the present model as well as in SLCM. This is lifted by the quantum fluctuation and the N$\rm \acute e$el-type long-range order of $T^z $ is realized at $J=1$ and $T=0$. 
In contrast to a number of studies at $T=0$, little is known about the finite-$T$ quantum states in SLCM. One of the reason is that any ordered phases are not expected to exist at finite temperature except for $J_z=J_x$.\cite{tanaka_compass} There is an ordered phase in the present model, and the finite temperature phase diagram is obtained by the QMC simulation. 

By utilizing several methods, as well as QMC, we conclude that there is a tricritical point around $J=1$ at finite temperature. In the scheme of the 1D+MF method, this is understood in analogy with the magnetization curve in the transverse-Ising model, and originates from the directional depending interaction. The present results provide clue information to reveal finite $T$ quantum states in other-types of the compass models.
It is also shown that, even on the geometrical frustrated lattice, a conventional Ising model does not show a tricritical point. As an example, let us consider the Ising model on a checkerboard lattice where the interactions along the horizontal/vertical and diagonal directions are of $J_z  S^z_i S^z_j$ and $J_x S^z_i S^z_j$, respectively. There is a critical point for a certain value of $J_x/J_z$, termed $J_c$, where a macroscopic number of degeneracy exists in the classical ground state. When the 1D+MF method is applied to this model, a self-consistent equation corresponding to Eq.~(\ref{eq:34}) is obtained as a conventional MF type of $\langle S^z \rangle \sim\tanh[\langle S^z \rangle /T]$. It is trivial that there is not a reflection point in this curve, in contrast to to Eq.~(\ref{eq:34}) and a tricritical point is not expected. 

In summary, we study the orbital compass model on a checkerboard lattice where the interactions along the horizontal/vertical and diagonal directions are given as $J_z T^z_i T^z_j$ and $J_x T^x_i T^x_j$, respectively. The classical and quantum models are analyzed by several analytical and numerical methods. We obtain the finite temperature phase diagram as a function of a ratio of $J=J_x/(2J_z)$. The N$\rm \acute e$el-type long-range ordr for $T^z$ occurs for $J \lesssim 1$, and a crossover from a disordered state to the $T^x$ correlated state along the diagonal chains is observed for $J \gtrsim 1$. A reentrant feature of the N$\rm \acute e$el-type $T^z$ order is shown around $J=1$ due to the thermal order-by-fluctuation mechanism. A tricritical point around $J=1$ is identified by QMC and 1D+MF methods. This is understood from a view point of the magnetization curve in the transverse Ising model and originates from the different types of the two competing interactions, i.e. $J_z T^z_i T^z_j$ and $J_x T^x_i T^x_j$. Excitation dynamics in this model are also examined. In the vicinity of the phase boundary in the N$\rm \acute e$el-type $T^z$ ordered state, the softening of the lowest coherent excitation peaks are confirmed along the lines of $k_x \pm k_z =\pm \pi$. This is interpreted as a precursor of the one-dimensional $T^x$ order stabilized above $J \sim 1.35$. The present studies do not only provide new insights in a combination of orbital frustration and geometrical frustration, but also help to reveal the finite $T$ quantum states in other-types of orbital compass models.

Authors would like to thank M.~Matsutomo and J.~Otsuki for the valuable discussions. This work was supported by KAKENHI from MEXT, Tohoku University ``Evolution'' program, and Grand Challenges in Next-Generation Integrated Nanoscience. JN is supported by the global COE program ``Weaving Science Web beyond Particle-Matter Hierarchy'' of MEXT, Japan. Parts of the numerical calculations are performed in the supercomputing systems in ISSP,  the University of Tokyo, and Kyoto University.

%*****************************************************************************

%*****************************************************************************


\begin{thebibliography}{99} 

%\begin{references}
\bibitem{book}
%\red{Full authors should be written in the reference.}
S.~Maekawa, T.~Tohyama, S.~E.~Barnes, S.~Ishihara, W.~Koshibae, and G.~Khaliullin, 
{\it Physics of Transition Metal Oxides}, 
(Springer Verlag, Berlin, 2004), and references therein. 

\bibitem{khomskii03}
D.~I.~Khomskii, and M.~V.~Mostovoy,
\journal{J. Phys. A: Math. Gen.}{36}{9197}{2003}.

\bibitem{Ishihara97}
S.~Ishihara, M.~Yamanaka, and N~Nagaosa,
\journal{\PRB}{56}{686}{1997}.


\bibitem{Kubo}
K.~Kubo, 
\journal{\JPSJ}{71}{1308}{2002}.

\bibitem{Nussinov_EPL}
Z.~Nussinov, M.~Biskup, L.~Chayes, and J.~van~den~Brink, 
\journal{\EPL}{67}{990}{2004}.

\bibitem{Khaliullin_Okamoto}
G.~Khaliullin and S.~Okamoto,
\journal{\PRB}{68}{205109}{2003}.

\bibitem{Feiner}
L.~F.~Feiner, A.~M.~Ole\'s, and J.~Zaanen,
\journal{\PRL}{78}{2799}{1997}.



\bibitem{Khaliullin_Maekawa}
G.~Khaliullin and S.~Maekawa,
\journal{\PRL}{85}{3950}{2000}.

\bibitem{Kugel}
K.~I.~Kugel, and D.~I.~Khomskii, 
\journal{Sov. Phys. Usp.}{25}{231}{1982}. 

\bibitem{Orus}
R.~Or\'us, A.~C.~Doherty, and G.~Vidal,
\journal{\PRL}{102}{077203}{2009}.

\bibitem{Nussinov}
Z.~Nussinov and G.~Ortiz,
\journal{\PRB}{77}{064302}{2008}.


\bibitem{Brz07}
W.~Brzezicki, and A.~M.~Ole\'s,
\journal{\PRB}{82}{060401(R)}{2010}.

\bibitem{Doucot}
B.~Dou\c{c}ot, M.V. Feigel'man, L.B. Ioffe, and A. S. Ioselevich,
\journal{\PRB}{71}{024505}{2005}.

\bibitem{Gladchenko}
S.~Gladchenko, D.~Olaya, Eva Dupont-Ferrier, B.~Dou\c{c}ot, L.~B.~Ioffe, and M.~E.~Gershenson,
\journal{\NATP}{5}{48}{2009}.

% \bibitem{kanamori}
% J.~Kanamori,
% \journal{\PTP}{17}{177}{1957}, \journal{\PTP}{17}{197}{1957}.



\bibitem{Jackeli}
G.~Jackeli and G.~Khaliullin,
\journal{\PRL}{102}{017205}{2009}.


\bibitem{katsufuji}
Y.~Horibe, M.~Shingu, K.~Kurushima, H.~Ishibashi, N.~Ikeda, K.~Kato, Y.~Motome, N.~Furukawa, S.~Mori, and T.~Katsufuji,
\journal{\PRL}{96}{086406}{2006}.

\bibitem{motome04}
Y.~Motome, and H.~Tsunetsugu,
\journal{\PRB}{70}{184427}{2004}.

\bibitem{vernay04}
F.~Vernay, K.~Penc, P.~Fazekas, and F.~Mila,
\journal{\PRB}{70}{014428}{2004}.

\bibitem{tomiyasu}
K.~Tomiyasu, M.~K.~Crawford, D.~T.~Adroja, P.~Manuel, A.~Tominaga, S.~Hara, H.~Sato, T.~Watanabe, S.~I.~Ikeda, J.~W.~Lynn, K.~Iwasa, and K.~Yamada,
\journal{\PRB}{84}{054405}{2011}.

\bibitem{Clay10}
R.~T.~Clay, H.~Li, S.~Sarkar, S.~Mazumdar, and T.~Saha-Dasgupta,
\journal{\PRB}{82}{035108}{2010}.


% \bibitem{Diaz}
% S.~Diaz, S.~de~Brion, G.~Chouteau, B.~Canals, V.~Simonet, and P.~Strobel,
% \journal{\PRB}{74}{092404}{2006}.


% \bibitem{Watanabe}
% T.~Watanabe, S.~Hara, and S.~Ikeda,
% \journal{\PRB}{78}{094420}{2008}.

\bibitem{Batista_Nussinov}
C.~D.~Batista and Z.~Nussinov,
\journal{\PRB}{72}{045137}{2005}.



%%%%%%%%%%%%%%%%%%%

\bibitem{WL}
F.~G.~Wang and D.~P.~Landau,
\journal{\PRL}{86}{2050}{2001}.

\bibitem{chen07}
H.-D.~Chen, C.~Fang, J.~Hu, and H.~Yao,
\journal{\PRB}{75}{144401}{2007}.



\bibitem{ALPS}
A.~F.~Albuquerque, F.~Alet, P.~Corboz, P.~Dayal, A.~Feiguin, L.~Gamper, E.~Gull, S.~G\"urtler, A.~Honecker, R.~Igarashi, M.~K\"orner, A.~Kozhevnikov, A.~L\"auchli, S.~R.~Manmana, M.~Matsumoto, I.~P.~McCulloch, F.~Michel, R.~M.~Noack, G.~Pawlowski, L.~Pollet, T.~Pruschke, U.~Schollw\"ock, S.~Todo, S.~Trebst, M.~Troyer, P.~Werner, S.~Wessel,
\journal{\JMMM}{310}{1187}{2007}.

\bibitem{Todo}
S.~Todo and K.~Kato,
\journal{\PRL}{87}{047203}{2001}.


\bibitem{Schmid}
G~.Schmid, S.~Todo, M.~Troyer, and A.~Dorneich,
\journal{\PRL}{88}{167208}{2002}.


\bibitem{Evertz}
H.~G.~Evertz
\journal{Adv.~Phys.}{52}{1}{2003}.

\bibitem{Friedan}
D.~Friedan and Z.~Qiu, and S.~Shenker,
\journal{\PRL}{52}{1575}{1984}.

\bibitem{Landau81}
D.~P.~Landau, and R.~H.~Swendsen,
\journal{\PRL}{46}{1437}{1981}.

\bibitem{tanaka_compass}
T.~Tanaka and S.~Ishihara,
\journal{\PRL}{98}{256402}{2007}.


% \bibitem{Henley}
% Christopher~L.~Henley,
% \journal{\PRL}{96}{047201}{2006}.

% \bibitem{Naka}
% M.~Naka, A.~Nagano, and S.~Ishihara,
% \journal{\PRB}{77}{224441}{2008}.




%\end{references}
\end{thebibliography}
\end{document}